\documentclass[twocolumn,showpacs,preprintnumbers,amsmath,amssymb]{revtex4}
\usepackage{graphicx}

\begin{document}
\preprint{\vbox{\hbox{UCB-PTH-02/05}},
  \vbox{LBNL-49487}}
\title{Signatures of Baryogenesis in the MSSM}
\author{Hitoshi Murayama}
\author{Aaron Pierce}
\affiliation{
Department of Physics, 
University of California, 
Berkeley, CA~~94720, USA}
\affiliation{Theory Group, 
Lawrence Berkeley National Laboratory, 
Berkeley, CA~~94720, USA}
\date{\today}
\begin{abstract}
  We revisit the electroweak baryogenesis within the context of the
  minimal supersymmetric standard model (MSSM), studying its potential
  collider signatures.  We find that this mechanism of baryogenesis
  does not give a new CP violating signal at the $B$-factories.  The
  first circumstantial evidence may come from enhanced $B_s$ or $B_d$
  mixing.  If a light right-handed scalar top and Higgs boson are found as
  required, a linear collider represents the best possibility for
  confirming the scenario.
\end{abstract}
\maketitle
\setcounter{footnote}{0}
\setcounter{page}{1}
\setcounter{section}{0}
\setcounter{subsection}{0}
\setcounter{subsubsection}{0}

\section{Introduction}\label{sec:intro}

The Minimal Supersymmetric Standard Model (MSSM) contains multiple
CP-violating complex phases.  This is in marked contrast to the
Standard Model which has only one phase in the CKM matrix.  Since the
Standard Model does not provide sufficient CP violation to account for
the observed baryon asymmetry of our universe, this new contribution
to CP violation is welcomed.

However, even with the additional sources of CP violation available in
the MSSM, it is non-trivial to achieve a sufficient baryon asymmetry.
Numerous groups have made detailed quantitative analyses of the
asymmetry.  For example, see
\cite{Carena:2000id,Carena:1997gx,Quiros:2000wk,Cline,Huet:1995sh}.
These studies have placed stringent constraints on the allowed
parameters of the MSSM.  Data from LEP further eliminate a large part
of this parameter space.

The starting point for our analysis is the constrained region of
parameter space that satisfies bounds from LEP and produces a
sufficient baryon to photon ratio.  We discuss the allowed parameter
space in section \ref{sec:status}.  Assuming that the MSSM
baryogenesis scenario is correct, and we lie in this region of
parameter space, we investigate the consequences that would be
accessible in collider physics.

Because $B$-physics has been viewed as a potential proving ground for
theories of baryogenesis, we pay particular attention to the
$B$-physics consequences of the scenario, addressing them in sections
\ref{sec:bsg} and \ref{sec:bmixing}.  We find an overall enhancement
in $B_s$ and $B_d$ mixing to be the signature of baryogenesis in $B$
physics.  However, the effect is rather subtle, with no new effects in
CP violation contrary to naive expectations.  The observation of light
scalar top quark, charginos and Higgs boson are necessary conditions
to confirm the scenario.  However, the mere observation will not
provide the proof of new CP violating phases necessary for the
baryogenesis.  We find that a linear collider represents the best tool
for determining whether a complex phase in the MSSM is responsible for
the cosmological baryon asymmetry.

\section{Status of the MSSM Baryon Asymmetry}\label{sec:status}

In this section, we discuss the MSSM spectrum that is selected by the
constraint that the MSSM provides a large enough baryon asymmetry.
The literature \cite{Cline,Carena:2000id,Quiros:2000wk} is in good
agreement on qualitative features of the spectrum selected out relevant
to collider signatures.  While there is not a complete quantitative
agreement on the amount of baryon asymmetry that can be generated in
this scenario, recent calculations do agree within an order of
magnitude, which is impressive considering that they take very
different approaches in calculating the baryon asymmetry.

The crucial CP violation responsible for the baryon asymmetry is in the
chargino sector.  If we write the chargino mass matrix as
\begin{equation}\label{eqn:massmatrix}
  {\cal M}_{C}=\left(
    \begin{array}{cc}
      M_{2} & \sqrt{2} M_{W} \sin \beta \\
      \sqrt{2} M_{W} \cos \beta & \mu e^{i \phi_{\mu}}
    \end{array} \right),
\end{equation}
the complex phase $\phi_{\mu}$ represents a source of CP violation not
present in the standard model. As explained in \cite{Cline:2000fh},
this phase leads to the dominant contribution to the baryon asymmetry.
As the universe undergoes a first-order electroweak phase transition,
bubbles of the true vacuum (where the gauge bosons are massive)
nucleate.  Charginos in the unbroken phase can then scatter off the
expanding bubble walls.  The complex phase gives rise to a classical
force that separates $\tilde{H}_{u}$ from $\tilde{H}_{d}$.  This, in
and of itself, does not create a baryon asymmetry.  However, the
asymmetry between higgsinos can be transformed into a chiral quark
asymmetry through higgsino scatterings off gluinos or stops.  Then
the chiral quark asymmetry can be further transformed into a baryon
asymmetry through the electroweak anomaly, which only acts on the
left-handed fields and violates $B+L$.

The qualitative picture outlined above only achieves quantitative
success for specific regions of SUSY parameter space.  One significant
constraint is that the phase transition must be first-order.  Without
a first-order transition, the bubble picture is not valid at all.  
Only after including two-loop corrections to the effective potential
does it become possible to achieve a first order phase 
transition \cite{Espinosa}.   
Moreover, the right-handed stop must be as light as
possible.  In particular, the right-handed stop mass should be less
than $m_{t}$ \cite{Cline:2000fh}.  While there have been searches for
the stop at both LEP and the Tevatron, the limits are somewhat model
dependent.  Limits from LEP indicate that the lightest stop is heavier
than roughly 90 GeV, while the limit from CDF indicates
$\tilde{m}_{t_{1}} >120$ GeV if the lightest neutralino is not too
heavy \cite{CDF,LEPStop}. In either case, the current limit is still
consistent with the requirement to have a stop light enough to achieve
a first order phase transition.

Electroweak baryogenesis also constrains $\tan \beta$.  If $\tan
\beta$ is too large, the CP asymmetry vanishes.  To see this, note
that as $\tan \beta \rightarrow \infty$, an entry in the chargino mass
matrix of Equation (\ref{eqn:massmatrix}) vanishes, and the phase of
$\mu$ can be rotated away with impunity.  Taking these considerations
into account, it is suggested in \cite{Cline} that a value of $\tan
\beta \approx 3$ is preferred.  The group of \cite{Carena:2000id} has
suggested that $\tan \beta < 6$ is necessary \cite{QuirosSays}.

Taking these values into account, on the other hand, it is not trivial
to avoid the bounds on the lightest Higgs mass from LEP II
\cite{LEPHiggs}.  At the tree-level, the lightest Higgs mass is
smaller for smaller $\tan\beta$, and needs to be boosted by the
radiative correction that goes approximately as
\begin{equation}\label{eqn:higgscorrection}
  \Delta m_{h}^{2} \simeq \frac{3}{4 \pi^{2}} \frac{m_{t}^{4}}{v^{2}} \log
  \left( \frac{\tilde{m}_{t_{1}} \tilde{m}_{t_{2}}}{m_{t}^{2}}\right),
\end{equation}
where $v \approx 174$ GeV.  Since $\tilde{t}_{R}$ must be light,
$\tilde{t}_{1}$ is also light, and the correction will be small unless
$\tilde{m}_{t_{2}}$ is somewhat sizeable.  To evade the Higgs mass
bound from LEP, we take $\tilde{m}_{t_{L}}$ to be 1 TeV.  According to
reference \cite{CarenaStop}, this is the minimum value necessary, and
even heavier values are necessary over most of the parameter space.   
Nevertheless, our conclusions regarding $B$ physics are unchanged for
the case of even heavier stops \footnote{Of course, the fine-tuning
  problem is exacerbated for larger values of the stop mass.}.

A non-zero $\phi_{\mu}$ generically has important consequences for
phenomenology.  In particular, one expects SUSY contribution to
electric dipole moments (EDM) to be too large.  This constrains the
masses of first two generations of superpartners.  Reference
\cite{EDM} finds that $\tilde{m}_{1,2} \sim$ 10 TeV is necessary to
avoid the EDM constraint for $\phi_\mu \sim O(1)$.  Because the baryon
asymmetry results from the chargino scattering at the time of phase
transition, charginos cannot be too heavy.  Reference \cite{Cline}
finds that $\mu \approx M_{2} < 250$ GeV to provide sufficient
asymmetry, but the chargino masses must also satisfy the LEP bound
$m_{\chi} >$ 103.5 GeV \cite{LEPChargino}.  For simplicity, we set the
$A$-terms to zero for quantitative analyses because it is not important.

This is an extremely well-defined scenario.  We investigate the
consequences of this scenario in the following sections.  It is
important to note that the phase $\phi_{\mu}$ is the key to
baryogenesis, and not the phase of $A_{t}$.  This causes a qualitative
change in the possible $B$-physics signatures.  Many previous
analyses \cite{Worah1,Worah2}, concentrated on the phase of the
SUSY-breaking parameter $A_{t}$ found in the stop mixing matrix.

\section{Radiative b Decay} \label{sec:bsg}

One place where one might expect evidence for this scenario to show up
is in the $b \rightarrow s \gamma$ decay.  Indeed, this turns out to
be a good place to put a constraint on the charged Higgs boson in the
MSSM, but not particularly useful to find evidence for or exclude the
scenario.

There is an extensive literature on the effects of supersymmetry on
this process
\cite{Bertolini,Barbieri:bsg,Chua:bsg,Baek:bsg,Goto:bsg,Aoki:bsg}.  A
notable result is that the branching ratio is always enhanced by the
charged Higgs boson exchange.  However, given the moderate values of
$\tan\beta$ required to achieve sufficient baryon asymmetry in the
scenario and LEP constraint on the lightest Higgs boson mass, the
charged Higgs boson is preferred to be heavy and its effect is
negligible.  On the other hand, if the lightest Higgs boson will not
be discovered soon, the requirement that sphaeleron erasure not wash
out the baryon asymmetry could force a somewhat lighter charged Higgs
boson mass \cite{Quiros:2000wk}.  It is still possible to have a
relatively light charged Higgs consistent with the LEP constraint, if
we allow multi-TeV left-handed stop. In this case, the tension with
the $b \rightarrow s \gamma$ branching ratio is exacerbated.  In our
numerical studies, we take the charged Higgs to be somewhat heavy, 1
TeV, to alleviate some of this tension with $b \rightarrow s \gamma$,
and assume that the lightest Higgs boson will be discovered soon.

\begin{figure}
  \includegraphics[width=\columnwidth]{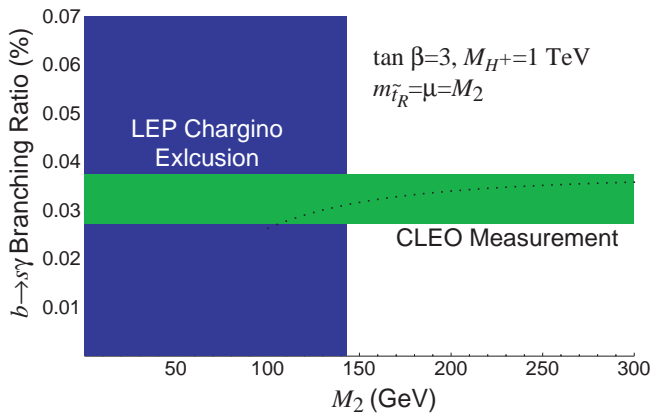}\label{fig:CLEO}
\caption{The $b \rightarrow$ $s \gamma$ branching ratio.  We have shown
  the one-sigma region for the CLEO measurements, combining all errors
  in quadrature.  Also shown is the region excluded by the LEP bound,
  $m_{\chi} > 103.5$ GeV.  We set $\tan \beta=3$, $\phi_{\mu}=\pi/2$,
  $M_{2}=\mu=\tilde{m}_{tR}$.  Other third generation sparticles are
  at a TeV, and the first and second superpartners are at 10 TeV.  We
  neglect the gluino contribution not required by baryogenesis, and 
  set the scalar trilinear coupling to zero.} 
\end{figure}

A contribution to $b \rightarrow$ $s \gamma$ that is necessarily there
in this scenario, independent of the charged Higgs mass, is the
chargino exchange.  As long as the left-handed stop is
sufficiently heavy \cite{Quiros:2000wk}, the phase of $\mu$ drops out,
and this diagram has a definite sign, opposite to the sign of the
Standard Model and charged Higgs contributions.  Therefore, the
chargino contribution tends to help agreement between the prediction
\cite{Kagan} and the branching ratio measured at CLEO \cite{CLEO}.  We
plot the branching ratio in Figure \ref{fig:CLEO}.  The message is
that this scenario is perfectly consistent with the $b \rightarrow$ $s
\gamma$ branching ratio constraint.

There is also a potentially important contribution from gluino
exchange if there is flavor mixing among the squarks.  This is a
contribution generically possible in the MSSM (currently with only
mild constraint; see \cite{Darwin}) and not required in the
baryogenesis.  Therefore, we do not regard the gluino contribution as
a signal of baryogenesis and neglect it in the rest of the paper.

As alluded to above, if the lightest Higgs boson is not discovered
soon, smaller values of the charged Higgs boson mass may be necessary.
However, the $ b \rightarrow$ $s \gamma$ branching ratio constrains
how light the charged Higgs boson can be.  In particular, to be
consistent with CDF bound on stop, assuming $\mu \approx M_{2}$, we
find that $m_{H^{\pm}} > 430$  GeV to have a small enough branching ratio
for $b \rightarrow s \gamma$.  To be consistent with the LEP bound on
stop (which is slightly less dependent on the neutralino masses) the
charged Higgs mass can be as light as 380 GeV.  To get a much lighter
value of the charged Higgs mass would require a model dependent gluino
contribution to cancel off some of the $b \rightarrow$ $s \gamma$
amplitude.

As we have already stated, $\phi_{\mu}$ decouples from the chargino
exchange diagram when there is large splitting between the stops.
This means that there is no chance to induce a large asymmetry ${\cal
  A}_{CP}$ for this decay.  Other studies \cite{Aoki:bsg,Baek:bsg}
have found potentially observable CP asymmetries, but they looked over
much larger regions of parameter space, and allowed a large phase for
$A_{t}$, which is not an essential feature of the MSSM baryogenesis
scenario. Using the formalism of \cite{Kagan}, we have investigated
the possibility of an an asymmetry ${\cal A}_{CP}$.  We find that in
the MSSM baryogenesis picture, where only the right-handed stop is
light, and the only complex phase is $\phi_{\mu}$, the asymmetry
${\cal A}_{CP}$ is less than 1\%.

\section{$B$ Mixing}\label{sec:bmixing}

Another potential arena for finding deviations from the Standard Model
is $B$ mixing measurements.  In particular, $B$ factories are making
precise measurements of $B_{d}$ mixing, while Run II at the Tevatron
should allow vastly increased sensitivity to $B_{s}$ mixing.  The main
difficulty with using these two measurements as a probe of new physics
lies with the large theoretical errors involved in their calculation.

The deviations in $B_{s}$ mixing in this scenario are shown in Figure
2.  We take $\tan \beta =3$. We set $M_{2}=\mu$ for
simplicity, but this is a good assumption---the region $M_{2} \approx
\mu$ is favored to get sufficient baryon asymmetry in any case.  In
the figure, we have set the charged Higgs mass to be relatively heavy,
at 1 TeV because large charged Higgs masses are favored, to increase
the the lightest Higgs mass.  This, in effect, isolates the chargino
contribution to the mixing.  We see that the maximal deviation is
about $30\%$.  The plot in the figure is for
$\phi_{\mu}=\frac{\pi}{2}$, but we have checked that as long as the
lightest chargino mass is kept fixed, there is not a great deal of
sensitivity to $\phi_\mu$.  This is due to the large splitting between
the $\tilde{t}_{L}$ and the $\tilde{t}_{R}$ states.

\begin{figure}
\includegraphics[width=\columnwidth]{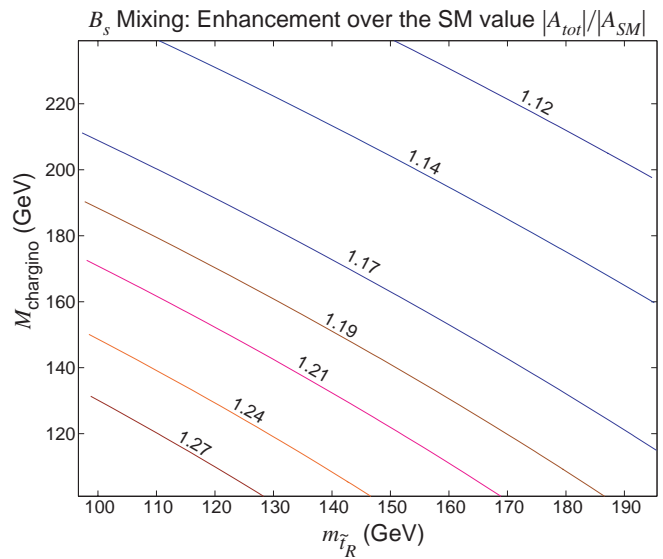}
\label{fig:Bmixing}
\caption{We show the enhancement of $B_{s}$ mixing in the MSSM
  baryogenesis scenario.  We have plotted contours of
  $|A_{tot}|/|A_{SM}|$, where $|A_{x}|$ is the magnitude of the
  amplitude for the $\Delta B=2$ mixing Hamiltonian in the full theory
  and Standard Model, respectively.  We take $\mu=M_{2}$, $\tan \beta
  =3$, $\phi_{\mu}=\pi / 2$, and $M_{H^{\pm}}=1$ TeV.} 
\end{figure}

The additional amplitude has the same phase as the Standard Model
contribution.  This is because the $\tilde{t}_R$ does not couple to
wino and hence its contribution singles out the higgsino coupling in
the chargino state.  Therefore the coupling is given by $h_t V_{ti}$
for $i=d,s,b$.  As a result, no significant deviation in CP-violating
observables induced by $B$-mixing is expected.  This is a direct
result of the Higgs mass bound, which forces the left-handed stop to
be heavy in the MSSM baryogenesis scenario.

We should note that there is a potential for gluino box diagrams to
contribute.  However, the point of this exercise is to capture the
essential signals of the baryogenesis scenario.  A large gluino
contribution is not essential in this framework, so we choose to
neglect these contributions.  Its presence may be an evidence for
supersymmetry, but not the MSSM baryogenesis.

Given the enhancement in $B$-mixing amplitudes in the scenario, the
problem is that current theoretical errors make it difficult to
identify this enhancement in $B$ mixing amplitudes.

There are two major sources of theoretical error for calculating
$B$-mixing in the Standard model.  First, there is imperfect knowledge
of the relevant CKM matrix elements.  In the Standard Model, $V_{td}$
is indirectly determined from $B$-mixing, which is not appropriate in
the presence of new physics contribution.  Other constraints on
$V_{td}$ come from $\epsilon_{K}$ and future measurements of $K
\rightarrow \pi \nu \overline{\nu}$.  However, these constraints will
remain relatively weak in the foreseeable future.  Another approach is
to assume unitarity of the CKM matrix.  In this case, the relatively
poor knowledge of $V_{ub}$ makes constraining $V_{td}$ difficult.  A
more precise determination of $V_{ub}$ (possibly down to 5--10\% level
\cite{Zoltan}) and $\sin 2\beta$ (to the few \% level at either LHC-B, 
B-TeV, or SuperBABAR \cite{SBabar}) in the future could determine $V_{td}$ 
to a sufficient accuracy.  On the other hand, $V_{ts}$, the relevant
parameter for $B_{s}$ mixing, is essentially determined by unitarity
$V_{ts} \approx - V_{cs} V_{cb}^*$ possibly down to 2--3\% level in
the future \cite{Zoltan}.

The second major source of theoretical error in $B$-mixing
calculations comes from hadronic matrix parameters which are found
from lattice calculations. $B_{d}$-mixing depends on the combination
$f_{B_{d}}^{2}B_{B_{d}}$, while $B_{s}$-mixing depends on
$f_{B_{s}}^{2} B_{B_{s}}$.  The latest unquenched evaluation from
JLQCD \cite{JLQCD}, quotes an error for $f_{B_{s}}^{2} B_{B_{s}}$ of
roughly $40 \%$.

Although the enhancement is not presently observable given current
theoretical errors, there is hope.  Over the next few years, it should
be possible to reduce the errors on $f_{B_{s}}^{2} B_{B_{s}}$ to
approximately 5\% in a quenched calculation in the next few years.
Similar errors for the unquenched calculation should be available on
the 5 to 10 year time scale \cite{AKornfeld}.  When such a calculation
is available, the enhancement of $B_{s}$-mixing could be discernible.

We also note that, as long as the model-dependent gluino contributions
are neglected, the ratio of $B_{s}$-mixing to $B_{d}$-mixing is
identical to the Standard Model prediction.  Because some lattice
uncertainties cancel in this ratio, it should be able to confirm this
prediction, particularly once advances are made in the determination
of $|V_{ub}|$.

Finally, we would like to mention that in the case where the charged
Higgs boson mass is lighter, it serves to enhance the $B$-mixing by a
few additional per cent at most.  The reason is that the biggest
enhancement to this process comes from the exchange of charginos and
the very light right-handed stop squark.

\section{MSSM Baryogenesis at a Linear Collider}\label{sec:LC} 

Although $B_{s}$ mixing may indirectly indicate light charginos and a
light stop, it will leave many questions unanswered.  First, is
$\phi_{\mu}$ non-zero?  Because it is this phase that provides the new
source of CP violation, we should not be convinced that the MSSM
explains baryogenesis unless we have produced evidence of a non-zero
$\phi_{\mu}$.

Obviously, it will be desirable to observe the light super-partners
directly.  In this scenario, only the right-handed stop and the
charginos may be observable at the Tevatron Run-II, LHC and the next
linear collider.  However, even if the light superpartners are
observed at the Tevatron or LHC, it seems unlikely that one could
discern that there is a non-zero phase for the $\mu$ parameter.

It seems that a linear collider, where one can make precision
measurements of the chargino system, is the best bet for observing
this phase.  It has been shown that a complete reconstruction of the
mass matrix (\ref{eqn:massmatrix}) is possible at a linear collider
where the charginos are accessible.  For example, see
\cite{Choi1,Choi2}.  To be explicit, the phase can be written in terms
of other observables as \cite{Choi1}:
\begin{eqnarray}
  \lefteqn{
    \cos \phi_{\mu}= 
    }
  \phantom{\frac{\Delta_{C}^{2}(2-c_{2L}^{2}-c_{2R}^{2})-8M_{W}^{2}\alpha_{C}}
    {\sqrt{(16 M_{W}^{2}-\Delta_{C}^{2}(c_{2L}-c_{2R})^{2})
        (4\alpha_{C}^{2}-\Delta_{C}^{2}(c_{2L}+c_{2R})^{2})}}}
  \nonumber \\
  \frac{\Delta_{C}^{2}(2-c_{2L}^{2}-c_{2R}^{2})-8M_{W}^{2}\alpha_{C}}
  {\sqrt{(16 M_{W}^{2}-\Delta_{C}^{2}(c_{2L}-c_{2R})^{2})
      (4\alpha_{C}^{2}-\Delta_{C}^{2}(c_{2L}+c_{2R})^{2})}},
\end{eqnarray}
where $\alpha_{C} \equiv (m_{\chi 2}^2+m_{\chi 1}^2-2 M_{W}^{2})$, and
$\Delta_{C} \equiv m_{\chi 2}^2-m_{\chi 1}^2$. $c_{2L} \equiv \cos{2
  \phi_{L}}$, and $c_{2R} \equiv \cos{2 \phi_{R}}$ are mixing angles
that arise in the diagonalization of the matrix of
Equation~(\ref{eqn:massmatrix}).  The angles $c_{2L}$ and $c_{2R}$ can
be extracted from measuring chargino production cross sections.  The
great precision to which a linear collider can measure the masses and
the mixings of the charginos allows access to the phase $\phi_{\mu}$,
proving the origin of the baryon asymmetry in the electroweak phase
transition. 

It would also be satisfying to observe this CP-violating phase through 
the measurement of a CP-odd quantity.  The prospects for such measurements 
at a linear collider have been discussed in \cite{Barger:2001nu}.

\section{Conclusions} \label{sec:conclusion}

While there are any number of ways to disprove the MSSM baryogenesis
scenario, it seems a non-trivial task to verify it.  Circumstantial
evidence could be found for the correct spectrum through measurements
of $B_{s}$ mixing, which could indicate the presence of a light stop
and light charginos.  The absence of a light stop or a fairly light
Higgs would immediately spell trouble for this scenario, and if they
are not found at the Tevatron or the LHC, it seems reasonable to
dismiss the MSSM as the source of baryon number in our universe.
However, even if these particles are present, we should not be
convinced that the MSSM provides the baryogenesis mechanism unless we
have evidence of a new CP violating phases. It is unlikely to observe
this phase by looking for CP violation at the $B$-factories.  To
ultimately determine that a phase is present in the chargino matrix
would fall upon a future linear collider.

\acknowledgments{ This work was supported in part by the DOE Contract
  DE-AC03-76SF00098 and in part by the NSF grant PHY-0098840. AP and
  HM would like to thank Z.~Ligeti, B.~Cahn, M.~Carena, M.~Quiros,
  and C.~Wagner for helpful conversations.}

\end{document}